\documentclass{article}

\usepackage{arxiv}

\usepackage[utf8]{inputenc} 
\usepackage[T1]{fontenc}    
\usepackage{hyperref}       
\usepackage{url}            
\usepackage{booktabs}       
\usepackage{amsfonts}       
\usepackage{nicefrac}       
\usepackage{microtype}      
\usepackage{lipsum}
\usepackage{graphicx}
\graphicspath{ {./images/} }

\title{CO.ME.T.A. - COvid-19 MEdia Textual Analysis. \\ A dashboard for media monitoring.}

\author{
 Emma Zavarrone \\
  IULM University\\
  \texttt{emma.zavarrone@iulm.it} \\
   \And
Maria Gabriella Grassia\\
 University of Naples Federico II\\
  \texttt{mgrassia@unina.it} \\
  \And
 Marina Marino\\
  University of Naples Federico II\\
  \texttt{ mari@unina.it} \\
 \AND
  Rosanna Cataldo \\
  University of Naples Federico II\\
 \texttt{rosanna.cataldo2@unina.it} \\
   \And
  Rocco Mazza \\
 University of Naples Federico II\\
   \texttt{ rocco.mazza@unina.it} \\
   \And
  Nicola Canestrari \\
  IULM University\\
 \texttt{nicolacanestrari.nc@gmail.com} \\
}

\begin{document}
\maketitle
\begin{abstract}
The focus of this paper is to trace how mass media, particularly newspapers, have addressed the issues about the containment of contagion or the explanation of epidemiological evolution. We propose an interactive dashboard: CO.ME.T.A.. During crises it is important to shape the best communication strategies in order to respond to critical situations. In this regard, it is important to monitor the information that mass media and social platforms convey. The dashboard allows to explore the mining of contents extracted and study the lexical structure that links the main discussion topics. The dashboard merges together four methods: text mining, sentiment analysis, textual network analysis and latent topic models. Results obtained on a subset of documents show not only a health-related semantic dimension, but it also extends to social-economic dimensions. 
\end{abstract}


\section{Introduction}
On Feb 11, 2020, WHO (World Health Organization) announced an official name for the syndrome coronavirus 2 (SARS-CoV-2), that is COVID-19. After a month the COVID-19 has been declared as pandemic. From December 2019 to March 2020 the COVID-19 has spread throughout China and afterwards through Italy, claiming victims and contagions\footnote{https://who.sprinklr.com/}. The focus of this paper is to trace how the mass media, particularly information on newspapers, have addressed the issues about the containment of contagion or the explanation of epidemiological evolution. Syilvie Briand, WHO general social media manager, affirms: "We know that every outbreak will be accompanied by a kind of tsunami of information, but also within this information you always have misinformation, rumors, etc. We know that even in the Middle Ages there was this phenomenon" (Zaracostats, 2020). Communication has an important role in the diffusion of behaviour and contagion, especially regarding the spreading of misinformation. During crises it is essential to spot the best communication strategies in order to respond to critical situations. Jin, Pang and Cameron’s (2007) studies underline how important it is to understand public’s emotional responses to crisis communication, in organizational and brand crisis but also in public and social crisis, such as infectious disease outbreaks (IDO) (Vijaykumar, Jin and Nowak, 2015). It is not an easy task to understand how communication from public health authorities or social media contents affect public attention and health-related risk evaluation and perception in these situations. It is crucial to constantly monitor public communication activities to find media response during the spread of a disease. To this end, it is important to monitor the information that the mass media and social platforms convey. Notable examples are Sharma et al. (2020), a Twitter based dashboard for analysing the COVID-19 misinformation, and Cinelli et al. (2020), who studied engagement and interest in the COVID-19 topic. In this paper we provide an in-depth textual comparison among established Italian and English newspapers from the end of January through an interactive dashboard. CO.ME.T.A. is the proposed shiny dashboard\footnote{https://rccmazza.shinyapps.io/cometa}, to represent an alternative way for reading the mass media perspective on tragic events about the viral infection. This contribution was made with the collaboration of CECOMS\footnote{https://www.iulm.it/it/ricerca/centri-di-ricerca/Cecoms} (Center for Strategic Communication Iulm University), in order to studies on the planning and design of strategic communication. The contribution at the state of the art is a tool for media monitoring during COVID-19 pandemic and a new media studies prospective for the study of crisis management. This paper is structured as follows: section 2 illustrates the methods and the data visualization tools used, while section 3 explains the corpus buildings procedures. Section 4 more specifically discusses the results for a source (\textit{The Guardian}), and section 5 presents future works. 
\begin{figure}
  \centering
   \includegraphics[scale=.30]{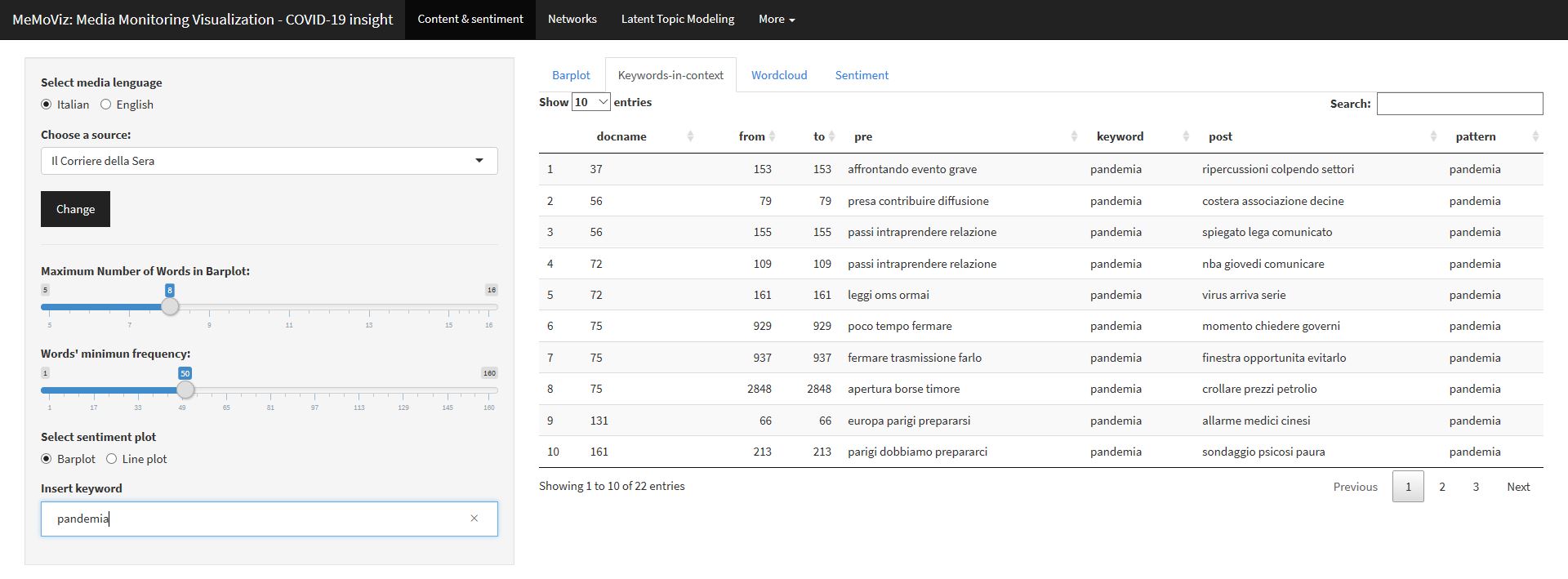}
  \caption{\textit{CO.ME.T.A.} User Interface}
  \label{fig:fig1}
\end{figure}

\section{Methodological features}
\label{sec:others}
CO.ME.T.A. is optimized to allow a friendly use even to those users who don’t have confidence with data analysis. The intuitive layout of user interface is divided between control panel on the left, plotting space on the right and menu bar with the methods on the upper side. The dashboard mixes four methods: text mining, sentiment analysis, textual network analysis and latent topic models. As concerns the latter model we propose a new visualization approach based on network to represent topics and words. Figure 2 shows the dashboard’s flowchart: (1) Content extraction and corpus pre-processing; (2) Sentiment analysis and descriptive study of texts: most frequent words and co-occurrence network analysis; (3) Application of a model to extract and identify the latent topics within the contents collected; (4) Plot network to represent each topic and semantic relationships between the extracted topics and terms. In the first step we defined preprocessing procedure for multilingual sources, using as reference the work done within the European project \textit{"Positive Messengers"} \footnote{https://positivemessengers.net/en/library.html}. After pre-treatment phase, the dashboard generates the final Document-Term Matrix and cut sparse words. DTM allows to describe the corpus through common visualizations, such as barplot of most frequent words and wordcloud. The sentiment analysis is performed using a baseline dictionary. The sentiment polarity is plotted during time lapse of documents publication. In addition, the DTM can be read like an affiliation matrix to analyse the semantic relationships. Using a textual network approach, we built a co-occurrence network and proposed the calculation of centrality measure between words. The last method is Latent Dirichlet Allocation model (Blei et al., 2003; Griffiths and Steyvers, 2004). LDA method is used to extract latent topics and subsequently construct the terms-topics matrix. The model allows to infer the latent structure of topics from recreating the documents in the corpus. This is possible by considering iteratively the relative weight of the topic in the document and the word in the topic. At the base of the LDA we find these assumptions: a) the documents are represented as mixtures of topics, where a topic is a probability distribution over words, as a generative and Bayesian inferential model; b) the topics are partially hidden, latent more precisely, within the structure of the document (Steyvers and Griffiths, 2007). Extracted the latent topics, the dashboard selects 20 most associated terms for each topic and it constructs a terms-topics two-mode matrix. Starting from this matrix, a two-dimensional network is plotted.
\begin{figure}[]
  \centering
   \includegraphics[scale=.5]{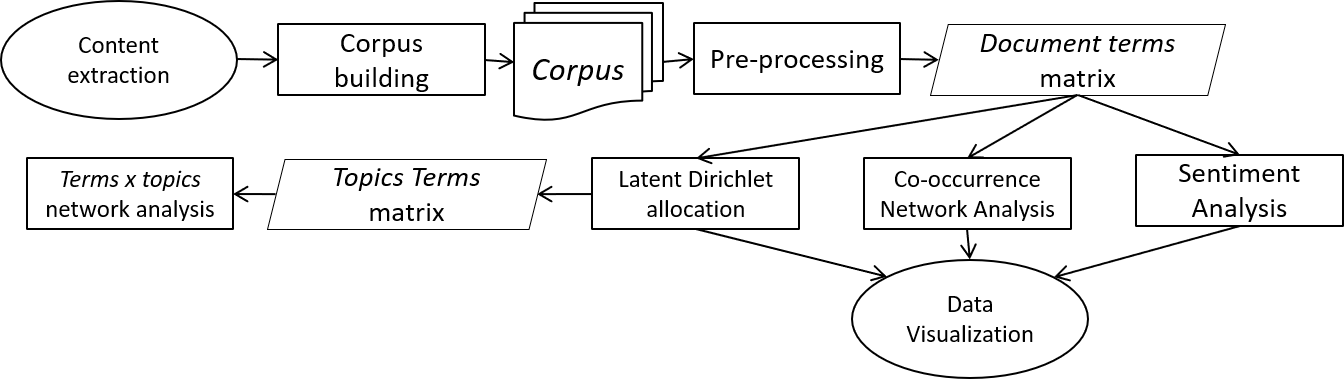}
  \caption{\textit{CO.ME.T.A.}'s flowchart}
  \label{fig:fig2}
\end{figure}

\section{Datasets Collection}
Textual datasets implemented in CO.ME.T.A. were built with a scraping of the on-line search results (the search key was “coronavirus”) of three Italian newspapers (“Il Corriere della Sera”, “La Repubblica”, “Il Sole 24 Ore”) and two English journals (“The New York Times”, “The Guardian”). We collect articles starting from 1 February, every 15 days there is an update. At the moment number of articles loaded in CO.ME.T.A. is 10328, 4380 in Italian language and 5940 in English language. 

\section{Dashboard results: The Guardian}
\label{sec:others}
This paragraph shows a concise and compact representation of analytical possibilities offered by the dashboard and an idea of the functions put in place for the users. Some of the results given by the main tools implemented in CO.ME.T.A. and related to The Guardian (collected from 2020-01-04 to 2020-03-11) are presented below, referred to as the first stage of alert, just before the declaration of pandemic status by the WHO. The wordcloud above shows not only a health-related semantic dimension, but it also extends to social-economic dimensions. A substantial prevalence of a negative sentiment is highlighted by the examination of the trend in a sentiment analysis on the documents. This underlines the high spikes occurred on January 25th, when the news reported first cases detected in the EU, on February 15th, when Chinese government implemented strict quarantine measures to contain the spreading of the virus from Hubei region, and on March 11th, which is the day of recognition of the disease as pandemic.
\begin{figure}
  \centering
  \includegraphics[scale=.45]{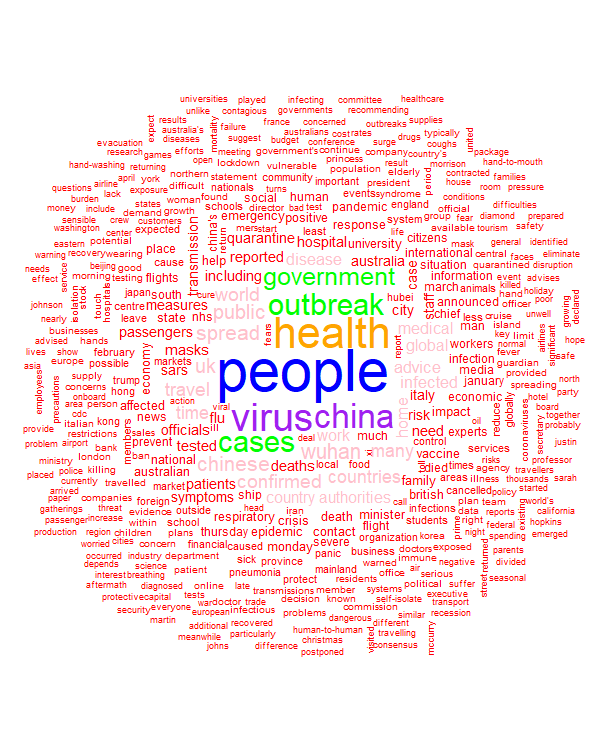}
  \caption{Wordcloud}
  \label{fig:fig1}
\end{figure}
\begin{figure}[!h]
  \centering
   \includegraphics[scale=.40]{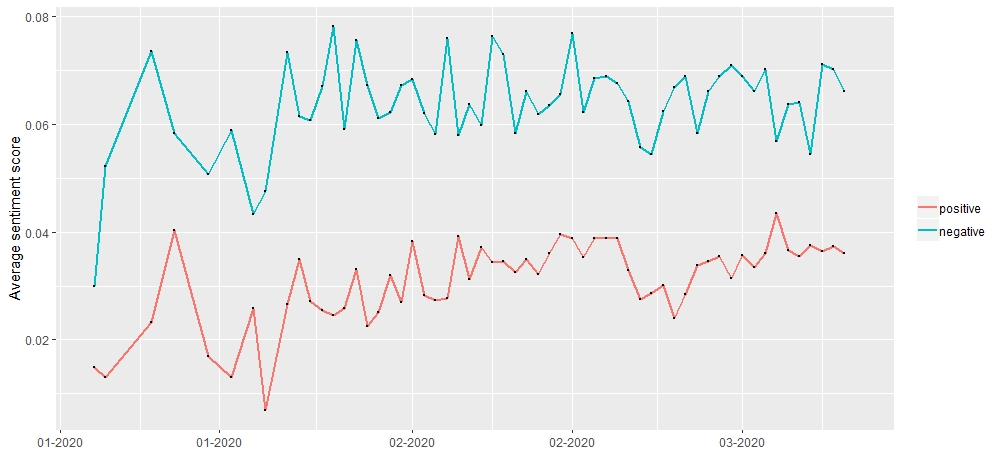}
  \caption{Sentiment line plot}
  \label{fig:fig2}
\end{figure}

With use of Latent Dirichlet Allocation 5 topics where extracted:
\begin{enumerate}
\item First topic is referred to medical and healthcare personnel and Personal Protecting Equipment (PPE);
\item Second topic is referred to political and economic impact that pandemic will have globally;
\item Third topic is referred to the spread of Covid-19 in China and containment measures taken by Beijing authorities;
\item Fourth topic is referred to SARS-CoV-2, the virus causing Covid-19, describing the virus and comparing the pathology with other diseases;
\item Fifth topic is referred to media and informative context, underling social response to the pandemic.
\end{enumerate}
\begin{table}
\caption{Topics extracted}
\centering
\begin{tabular}{lllll}
\toprule
Topic 1     & Topic 2     & Topic 3 & Topic 4 & Topic 5 \\
 \midrule
    people                                              & global                                              & cases                                               & people                                              & time                                                \\
health                                              & government                                          & virus                                               & virus                                               & people                                              \\
masks                                               & economy                                             & china                                               & outbreak                                               & public                                              \\
staff                                               & travel                                              & health                                              & flu                                              & disease                                            \\

    \bottomrule
  \end{tabular}
  \label{tab:table}
\end{table}
Through the words-topic network it is possible to observe how the terms are associated with the referred topic. The network is composed by latent topics, identified through the LDA technique and the words associated with the highest probability. This network allows to examine the links between these two dimensions, particularly howthe corpus are distributed among the topics. A node represents a term connected with different topics and indicates that it is not only present in both thematic groups, but it also represents a connection between semantic areas associated with each topic. Terms with higher degree centrality (Faust, 1997) are “people, virus, health, outbreak, china, public, uk, government, world, cases, wuhan, masks, staff, home, patients”. A high level of centrality in these terms means a strong attention to personal protective equipment and national health preparation to the crisis. Terms with high level of closeness centrality (Bonacich, 1991) are “outbreak, virus, china, government, world”. In this case the central semantic dimensions detected by the models are the outbreak of the pandemic and the global spreading of the disease. In the topic network it is possible to identify how the term “outbreak” links different topics related to semantic dimensions of economic, health and mediatic spheres.
\begin{figure}[!h]
  \centering
   \includegraphics[scale=.40]{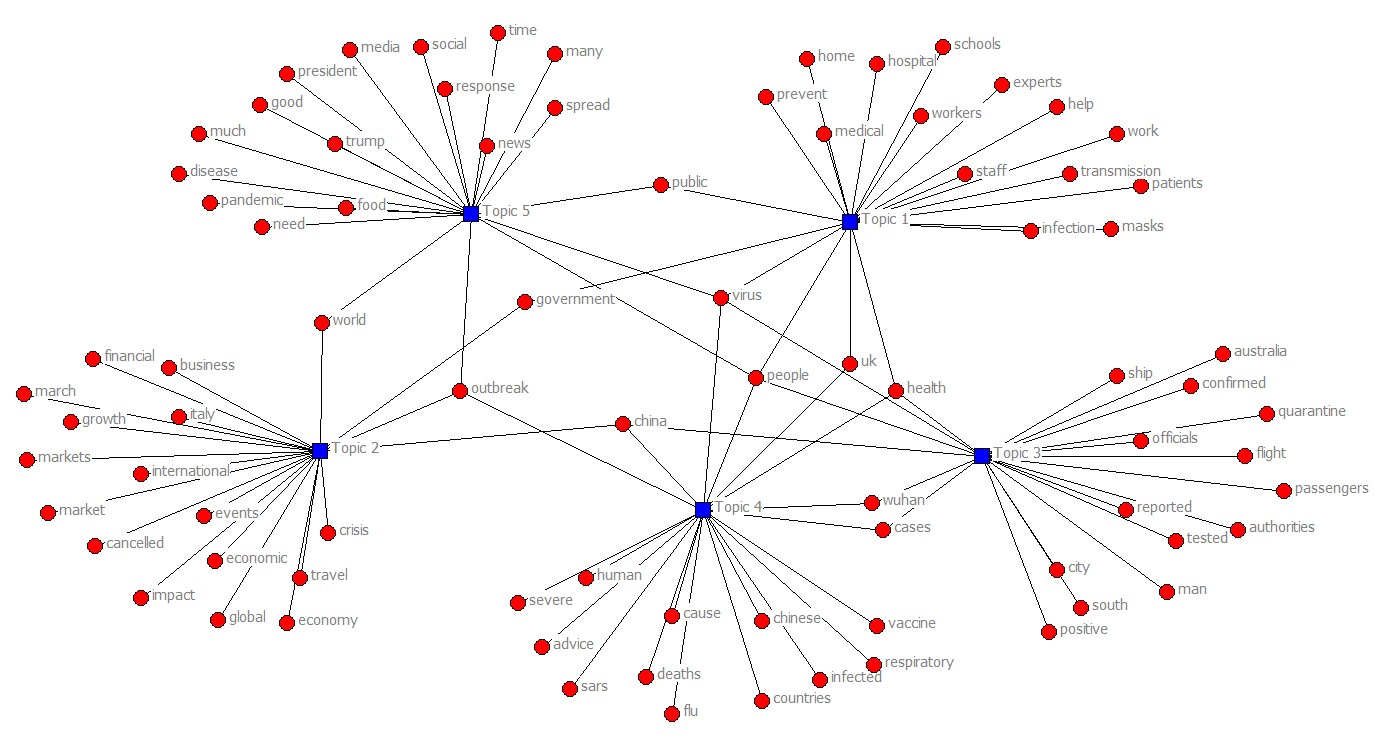}
  \caption{Topics-Terms network}
  \label{fig:fig2}
\end{figure}

\section{Future works}

Future works may take into consideration several directions, in order to optimize analysis of information and communication about COVID-19 spreading. Since the disease has spread globally, the intention of the research is to extend the datasets analysed to other important newspapers in other languages, such as Spanish or French. This aims to have a more complete and global representation of the response of mediatic communication to the virus. Another purpose of future researches is to identify a connection between sentiment trend extracted from articles and the epidemiological curve to quantify the effect given by death/contagious/healing rates to the communication. An implementation on the dashboard of a sentiment analysis on Twitter text from the community could give a description of the public feedback to news, giving indications to media to provide a better communication in crisis situations. One of the most interesting developments for future works is to identify a relation between sentiment given by user tweets and news articles compared with bulletins and speeches given by the Italian Prime Minister, Giuseppe Conte. Therefore, it could be defined if there is a correlation in the paired trends and extracting information about cause-effect phenomena. In future projects intent of the research is to implement the dashboard with two further analytical processes. Through Correspondence Analysis we aim to represent the association structure between a group of extracted keywords and analysed texts, to identify concepts directly unobservable but as results of the measurement of a group of variables. With application of neural networks, it will be possible to better classify texts through textual data measurements in content extraction and corpus pre-processing phases.

\begin{table}[!h]
\caption{Terms degree and closeness centrality values calculated on topic-terms matrix }
\centering
\begin{tabular}{llllllll}
\toprule
Terms    & Nor. Degree & Terms      & Nor. Degree & Terms    & Nor. Degree & Terms      & Closeness  \\
 \midrule
people   & 0,800       & public     & 0,400       & wuhan    & 0,400       & outbreak   & 0,537                \\
virus    & 0,800       & uk         & 0,400       & masks    & 0,200       & virus      & 0,535                \\
health   & 0,600       & government & 0,400       & staff    & 0,200       & china      & 0,535                \\
outbreak & 0,600       & world      & 0,400       & home     & 0,200       & government & 0,535                \\
china    & 0,600       & cases      & 0,400       & patients & 0,200       & world      & 0,535              \\
\end{tabular}
 \label{tab:table}
\end{table}

\bibliographystyle{unsrt}  


\end{document}